# Photon number resolution without optical mode multiplication

Anton N. Vetlugin,[1,2*] Filippo Martinelli,[1,2] Shuyu Dong,[1,3] and Cesare Soci[1,2,3*]

[1]*Centre for Disruptive Photonic Technologies, TPI, Nanyang Technological University, Singapore 637371*

[2]*Division of Physics and Applied Physics, School of Physical and Mathematical Sciences, Nanyang Technological University, Singapore 637371*

[3]*School of Electrical and Electronic Engineering, Nanyang Technological University, Singapore 639798*

*Corresponding authors. Email: a.vetlugin@ntu.edu.sg (A.N.V.); csoci@ntu.edu.sg (C.S.)

**Common methods to achieve photon number resolution rely on fast on-off single-photon detectors in conjunction with temporal or spatial mode multiplexing. Yet, these methods suffer from an inherent trade-off between the efficiency of photon number discrimination and photon detection rate. Here, we introduce a method of photon number resolving detection that overcomes these limitations by replacing mode multiplexing with coherent absorption of a single optical mode in a distributed detector array. Distributed coherent absorption ensures complete and uniform absorption of light among the constituent detectors, enabling fast and efficient photon number resolution. As a proof-of-concept, we consider the case of a distributed array of superconducting nanowire single-photon detectors with realistic parameters and show that deterministic absorption and arbitrarily high photon number discrimination efficiency can be achieved by increasing the number of detectors in the array. Photon number resolution without optical mode multiplication provides a simple yet effective method to discriminate an arbitrary number of photons in large arrays of on-off detectors or in smaller arrays of mode multiplexed detectors.**

## INTRODUCTION

Experimental quantum optics relies on detecting light of extremely low intensity – down to a single-photon level. Most of the existing single-photon detectors operate in the on-off regime, irrespective of the number of absorbed photons. At the same time, some applications such as linear optical quantum computation [1-3], quantum communication and key distribution [4], quantum light sources characterization [5], and quantum states preparation [6,7] require photon number resolving (PNR) detection [8-10]. Beyond this, PNR detectors would benefit several fields in the classical optics domain, including fluorescence-lifetime imaging microscopy [11], X-ray astronomy [12], lidars [13,14], elementary particle detection [15], and medical diagnostic [16-18].

Some types of photomultiplier tubes [19], single-photon avalanche diodes [20,21], visible light photon counters [22,23], and transition edge sensors [24] have intrinsic mechanisms that allow resolving the number of photons. Yet, intrinsic PNR detectors suffer from poor photon number resolution even at a few photon level (i.e., photomultiplier tubes, avalanche diodes), high dark count rate (i.e., visible light photon counters), slow operational rate (i.e., visible light photon counters, transition edge sensors), extreme regimes of operation (i.e., transition edge sensors) and high jitter time. Multiplexing is commonly used to overcome the limitations of intrinsic PNR detectors. The operation of multiplexed PNR detectors is based on splitting the incoming light into multiple temporal [25-27] or spatial [28-30] modes and detecting these modes independently by single-photon on-off detectors. This allows the exploitation of state-of-the-art detectors such as superconducting nanowire single-photon detectors (SNSPDs) [31-33], which are characterized by negligible dark count rate, high operational rate, and low jitter time simultaneously. Importantly, multiplexed PNR detectors are equivalent to intrinsic PNR detectors in photon number resolution if the number of multiplexed modes is large enough [34]. Moreover, multiplexing with intrinsic PNR detectors [6] further increases the maximum number of photons that can be resolved [35]. However, the multiplication of optical modes is associated with a decrease in temporal performance (temporal multiplexing), non-uniform illumination of the constituent detectors (spatial multiplexing), and additional losses (both of them), which is detrimental to practical PNR detection of light.

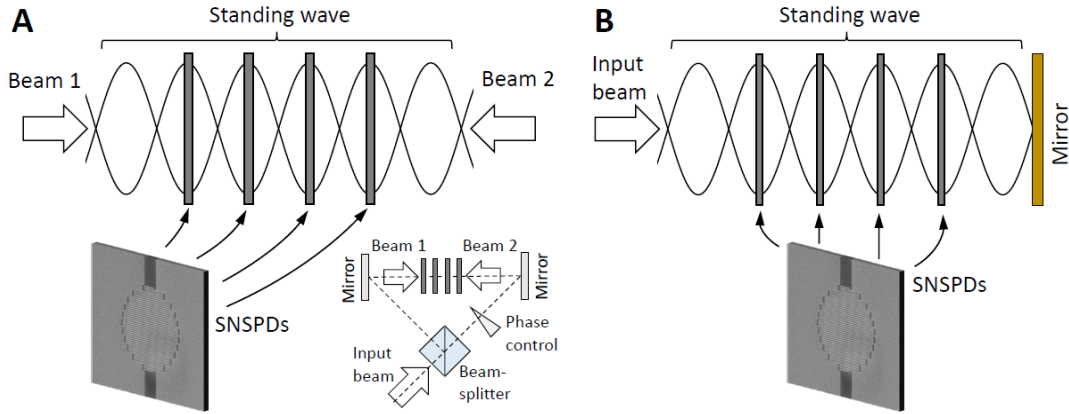

**Figure 1.** Coherent detection of light with a distributed detector. Multiple SNSPDs are placed at separate anti-nodes of the standing wave, guaranteeing total and uniform light absorption. A distributed detector can be fabricated in a phase-sensitive design (A) providing control over light absorption in a full range or in a phase-insensitive design (B) always operating in the total absorption regime. The inset in (A) shows an interferometric setup for a phase-sensitive detection. SEM images show NbTiN SNSPDs.

Here, we introduce a method of PNR detection without optical mode multiplication (Fig. 1). Unlike conventional schemes, our method relies on the coherent detection of light as a *standing* wave. We consider a distributed 1D arrangement of single-photon detectors where each detector is placed at a separate anti-node of the standing wave. This positioning of the detectors within a single optical mode guarantees their equal exposure and, consequently, uniform light absorption. We show how to tailor the optical response of constituent detectors so that the entire structure operates in the regime of total light absorption. This method can be considered as a generalization of the phenomenon of coherent perfect absorption with a single-layer absorber [36-51]. For the sake of the following discussion, we present the case of distributed detectors composed of SNSPDs, but the method can be immediately applied to the case of any type of distributed detector.

## RESULTS

### Salisbury screen design for a single-layer SNSPD

A typical bare meander nanowire used for fabricating SNSPDs absorbs light weakly and, thus, is unsuitable for efficient light detection. Even placing such an SNSPD in the Salisbury screen geometry, where a reflector – metallic mirror of Distributed Bragg Reflector (DBR) is placed beneath the SNSPD with a $\lambda/4$-spacer in between, Fig. 2A, does not allow to reach the total light absorption. We demonstrate this in Fig. 2, where we calculate the optical response of such a structure assuming a meander nanowire made of niobium-titanium nitride (NbTiN) film with a standard filling factor $f$=0.5 and incident light polarization parallel to the meander (grating) slits. In this case, the meander can be treated within the effective medium approximation as a uniform film with *effective* permittivity $\varepsilon_{\text{eff}} = \varepsilon_{\text{film}} f + \varepsilon_{\text{slit}}(1-f)$, where $\varepsilon_{\text{film}}$ and $\varepsilon_{\text{slit}}$ represent the permittivity of the superconductor nanowire film (NbTiN) and that of the nanowire surrounding region (dielectric matrix), respectively [52,53]. In Fig. 2B, we calculate how the intensity reflection $R$ and absorption $A=1-R$ coefficients (transmission is negligible) of the Salisbury screen structure at telecom wavelength of $\lambda = 1550$ nm change during deposition of the nanowire layer (solid line) on top of the spacer with the refractive index $n_{\text{sp}}$= 1.5. Reflection is suppressed and absorption reaches unity for the nanowire thickness of 15 nm. This is the point of maximum absorption of the SNSPD layer. Further increase of the SNSPD layer thickness induces a decrease in absorption. The circular diagram in Fig. 2C provides a useful representation of this behaviour. Here, the complex amplitude reflection $r$ coefficient of the structure is plotted for continuously increasing thickness of the SNSPD layer (representative thicknesses are indicated by symbols as in Fig. 2B). The thickness of the nanowire affects mostly the magnitude of $r$, which reduces to nearly zero for 15 nm thick nanowire layer. The reflection and absorption spectra of the Salisbury screen structure with the 15 nm thick nanowire are shown in Fig. 2D, where the optical constants for the reflector and spacer are assumed wavelength-independent while the dispersion of NbTiN film is accounted for. The complex relative permittivity of NbTiN film was taken from Ref. [54] where it was derived for a 5 nm thick film, the typical thickness of optimized SNSPDs (SEM images in Fig. 1). The characteristic oscillations seen in the optical response are related to the spacer thickness optimized for absorption at 1550 nm.

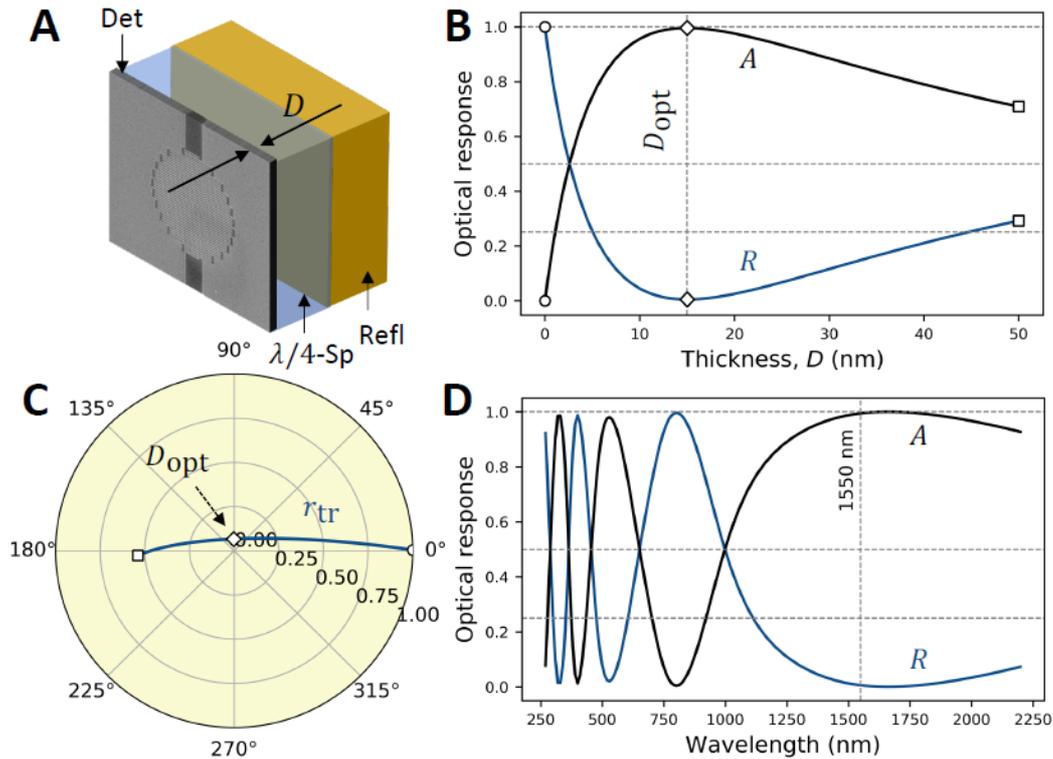

**Figure 2.** Salisbury screen SNSPD design. A: The nanowire detecting layer (Det) is separated from the reflector (Refl) by a $\lambda/4$-spacer, thus placing the nanowire at the anti-node of the standing wave. In the calculations, the mirror is treated as a semi-infinite layer with reflectivity of 99.9%. B: The optical response of the structure- reflection $R$ and absorption $A$, is calculated while sweeping the thickness $D$ of the nanowire. C: The corresponding amplitude reflection r coefficient is shown in the circular diagram, where the radial (angular) coordinate defines the magnitude (phase) of r. In B and C, empty circles and squares mark the beginning ($D=0$ nm) and end ($D=50$ nm) of the deposition, respectively, while diamonds mark the optimal thickness. D: The reflection $R$ and absorption $A$ spectra are calculated for an optimal thickness of $D_{opt}$ =15 nm. The nanowires are assumed to be carved out of a NbTiN film with permittivity $\varepsilon_{film} = (4.21 + i3.87\ )^2$ at 1550 nm.

Although full absorption could be achieved with a NbTiN thickness of 15 nm, practically the superconducting films need to be thinner than 10 nm, as the absorbed photon cannot break superconductivity in thicker layers and the internal efficiency drops. As a result, the Salisbury screen is replaced by a resonant cavity in detectors with high system efficiency, at the expense of sandwiching SNSPD between complex multi-layered structures or extreme phase sensitivity [55,56].

**Total absorption by partially absorbing nanowire layers**

The optical response of a bare nanowire meander is shown in Fig. 3. Assuming single side (traveling wave) illumination, as in Fig. 2, we calculate the intensity transmission $T_{tr}$ (red line), reflection $R_{tr}$ (blue line), and absorption $A_{tr} = 1 - T_{tr} - R_{tr}$ (solid black line) coefficients of the nanowire layer as a function of its thickness, Fig. 3B. All calculations are done using the transfer matrix method (see Methods below) within the effective index approximation (see Supplementary Materials for comparison with finite element method simulations). The optical response shown in Fig. 3B reveals that the maximum absorption is just 50%, which is a fundamental limit of traveling wave absorption of any subwavelength absorptive layer [57].

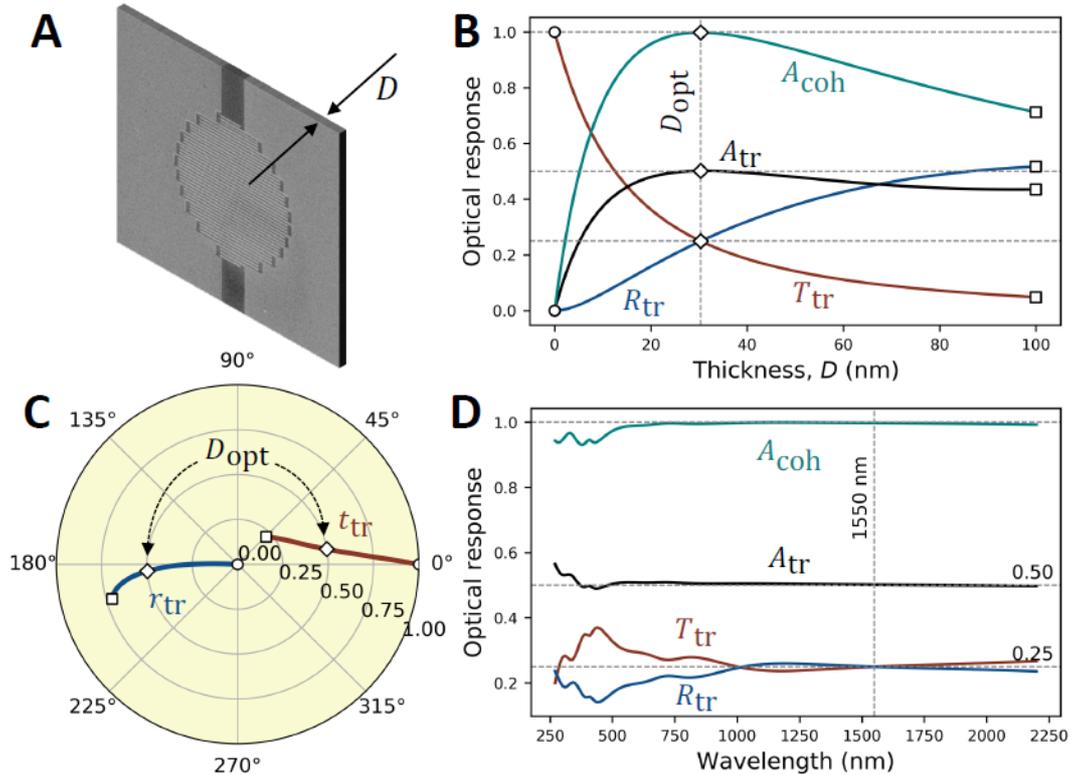

**Figure 3.** A single-layer NbTiN SNSPD absorber. A: Free-space nanowire absorber with variable thickness $D$. B: Optical response is evaluated under traveling (intensity transmission $T_{tr}$, reflection $R_{tr}$, and absorption $A_{tr}$ coefficients ) and standing (intensity absorption $A_{coh}$ coefficient) wave illumination as a function of the nanowire thickness $D$. C: The corresponding amplitude transmission $t_{tr}$ and reflection $r_{tr}$ coefficients are shown in the circular diagram, where the radial (angular) coordinate defines the magnitude (phase) of $t_{tr}$ and $r_{tr}$. In B and C, empty circles and squares mark the beginning ($D=0$ nm) and end ($D=100$ nm) of the deposition, respectively, while diamonds mark the optimal thickness. D: The transmission $T_{tr}$, reflection $R_{tr}$, and absorption $A_{tr}$ and $A_{coh}$ spectra are calculated for an optimal thickness of $D_{opt}$ =30 nm. Similar calculations for NbN and MoSi SNSPDs are provided in Supplementary Materials.

On the other hand, this fundamental limit can be overcome when the partially absorbing film is placed in a standing wave appearing as a result of interference of two counter-propagating coherent beams, as in Fig. 1A. When two beams are in phase, the anti-node (maximum electric field) of the standing wave coincides with the nanowire, resulting in enhanced absorption (coherent absorption regime) [37,42,48]. The coherent absorption coefficient, $A_{coh}$, of the nanowire for this regime is shown by the green line in Fig. 3B. Compared to traveling wave absorption, absorption of the standing wave doubles at the optimal thickness, $D_{opt}$ =30 nm, to reach 100%. The nature of this enhancement is clearly seen in the circular diagram in Fig. 3C. Here, the complex amplitude transmission $t_{tr}$ and reflection $r_{tr}$ coefficients are plotted for continuously increasing film thickness. Amplitude transmission and reflection coefficients are of equal amplitudes but opposite phases at the optimal film thickness (marked by diamonds): $t_{tr} \approx 0.5$ and $r_{tr} \approx -0.5$. The amplitudes of outgoing waves on both sides of the SNSPD are given by $t + r$ and thus cancel, resulting in perfect light absorption. Remarkably, such a mechanism is effective in an extremely broad range of wavelengths, resulting in coherent absorption greater than 93% throughout the entire spectral range and greater than 99% between 600 and 2200 nm, as shown for the film of optimal thickness in Fig. 3D. This is in stark contrast with the narrowband operation of the Salisbury screen due to the presence of a spacer. Conversely, when the input waves are out of phase, the amplitudes of the outgoing waves, given by $\pm(t - r)$, equal the amplitudes of the input waves and absorption is completely suppressed (coherent transmission regime) [37,48]. In this case, the nanowire absorber is positioned at the node of the standing wave, where the electric field vanishes.

The coherent absorption regime is equivalent to the Salisbury screen geometry. In the latter case, a standing wave results from interference between the incident and reflected on the reflector waves. Due to the $\lambda/4$-spacer and $\pi$-phase shift on the reflector, the nanowire is always placed at the anti-node. Thus, while the Salisbury screen and counter-propagating geometries provide the same maximum light absorption, the counter-propagating geometry provides the additional flexibility to control the absorption level between zero and unity by changing the mutual phase of two interfering beams. This may be of interest for feed-forward protocols or self-configuring optical networks [58,59].

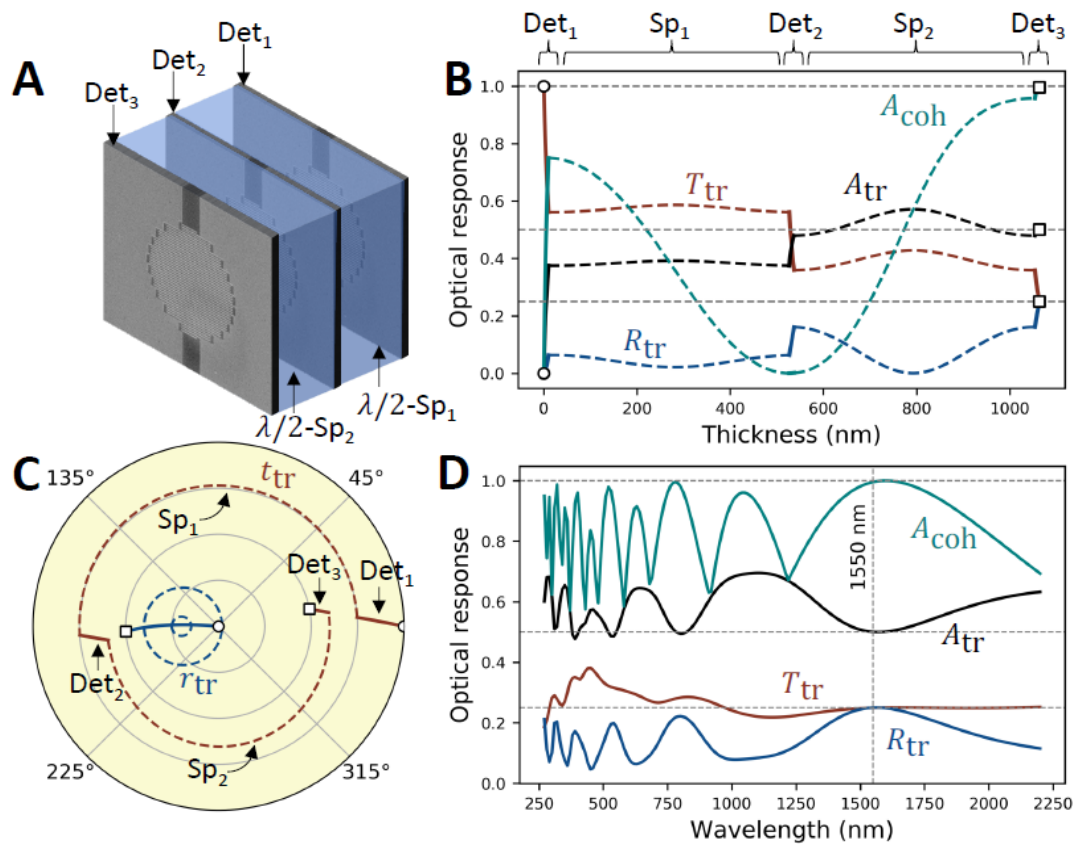

**Figure 4.** Three-layer distributed detector. A: Distributed detector consists of three nanowire layers – $Det_1$, $Det_2$ and $Det_3$ and two $\lambda/2$-spacers $Sp_1$ and $Sp_2$. B and C: the intensity (B) and amplitude (C) coefficients are evaluated continuously as layers are deposited, starting with $Det_1$. In (B) and (C), solid (dashed) lines correspond to the deposition of the nanowire layers (spacers). Empty circles and squares mark the beginning and end of the deposition. D: The spectrum of the entire structure.

The standing wave picture also suggests a simple way to achieve total light absorption from nanowire layers of arbitrary small thickness. Let us assume the nanowire layer of optimal thickness $D_{opt}$=30 nm is "cut" into multiple sublayers. The sublayers positioned in the same anti-node of the standing wave would interact equally with the electric field (unlike the case of a thick absorber stretching across different regions of the standing wave). Equivalently, each sublayer may be positioned at different anti-nodes of the standing wave while experiencing the same interaction with the electric field. Thus, the optical response of this distributed structure would be identical to that of the original single-layer nanowire absorber. For example, let us consider a distributed structure consisting of three sublayers (three SNSPDs), Fig. 4A, of thickness $D_{\text{opt}}/3 = 10$ nm each. For practical implementation, let us also assume that the SNSPDs are separated by dielectric spacers of thickness $d_{sp} = \lambda/(2n_{sp})$ rather than free space. Such spacers are known in thin film optics as "absentee" layers, as they do not change the reflection of the structure beneath [60]. Fig. 4B shows how the optical response of the structure changes as the three nanowire sublayers ($Det_1$, $Det_2$, $Det_3$ – solid lines) and the two dielectric spacers ($Sp_1$, $Sp_2$ – dashed lines) are deposited consecutively. The completed structure is characterized by the same optical response as the original single-layer SNSPD of optimal 30 nm thickness (cf. Fig. 3B).

    The contribution of each layer of the distributed absorber can be understood from the circular diagram in Fig. 4C. While the nanowire sublayers change the magnitudes of $t_{\text{tr}}$ and $r_{\text{tr}}$, each spacer induces a $\pi$-phase shift for transmission and a $2\pi$-phase shift for reflection, without changing the transmission and reflection magnitudes. Thus, the response of the entire multilayer structure is $t_{\text{tr}} \approx 0.5$ and $r_{\text{tr}} \approx -0.5$, as required to achieve coherent perfect absorption. Unlike coherent perfect absorption in the single nanowire layer with optimal thickness, however, the optical response of the multilayer structure strongly depends on wavelength, Fig. 4D. Such dependence becomes more pronounced as the number of layers increases, an issue that in thin film optics is commonly addressed by introducing achromatizing layers [60].

    Overall, distributed absorbers provide an alternative architecture to achieve total absorption in SNSPDs without the need for complex DBR or optical cavity structures that are fairly intolerant to the thickness of their components [61]. Furthermore, when implemented with independent on-off or PNR detectors, distributed coherent perfect absorption allows photon number resolution methods without spatial or temporal multiplexing.

**Absorption uniformity across constituent detectors**

Let us consider an array of equally spaced nanowire sublayers placed at the anti-nodes of a standing wave, Fig. 1A. In the regime of coherent perfect absorption, this arrangement allows complete absorption of multiple photons distributed across the SNSPD array, which may be employed for photon number resolution.

    For accurate photon number resolution, the number of constituent detectors in the array should be large enough to reduce the probability of absorbing more than one photon in the same constituent detector. For instance, to resolve a two-photon state with a probability of 90%, one would need at least 10 detectors (see discussion below). Thus, one should maximize the number of sublayers into which the optimal absorber thickness, $D_{\text{opt}}$, that guarantees coherent perfect absorption is split. Practically, this can be achieved by reducing the thickness of each nanowire sublayer (although, practically, it is challenging to fabricate high quality superconducting films thinner than 5 nm), or by decreasing the filling factor of the nanowire meanders in each sublayer absorber. In Fig. 5A, the optimal thickness is plotted as a function of the filling factor $f$ (solid blue line). Empty circles mark $D_{\text{opt}} = dN_{\text{Det}}$, where d=5 nm is the thickness of each nanowire sublayer and $N_{\text{Det}}$=5, 10, 15 is the number of layers (detectors) into which $D_{\text{opt}}$ can be "cut". The horizontal coordinates of these points (dashed vertical lines) define the filling factor of the nanowire meanders. On the same graph, we plot the absorption of a single bare nanowire sublayer (5 nm thick) under traveling wave illumination as a function of the filling factor (dashed black line). For instance, a distributed detector can be built out of five, ten or fifteen nanowire sublayers with a thickness of 5 nm each and a filling factor of 0.61, 0.30 and 0.20, respectively. A single bare nanowire with these parameters would absorb just 28%, 17% and 12% of light.

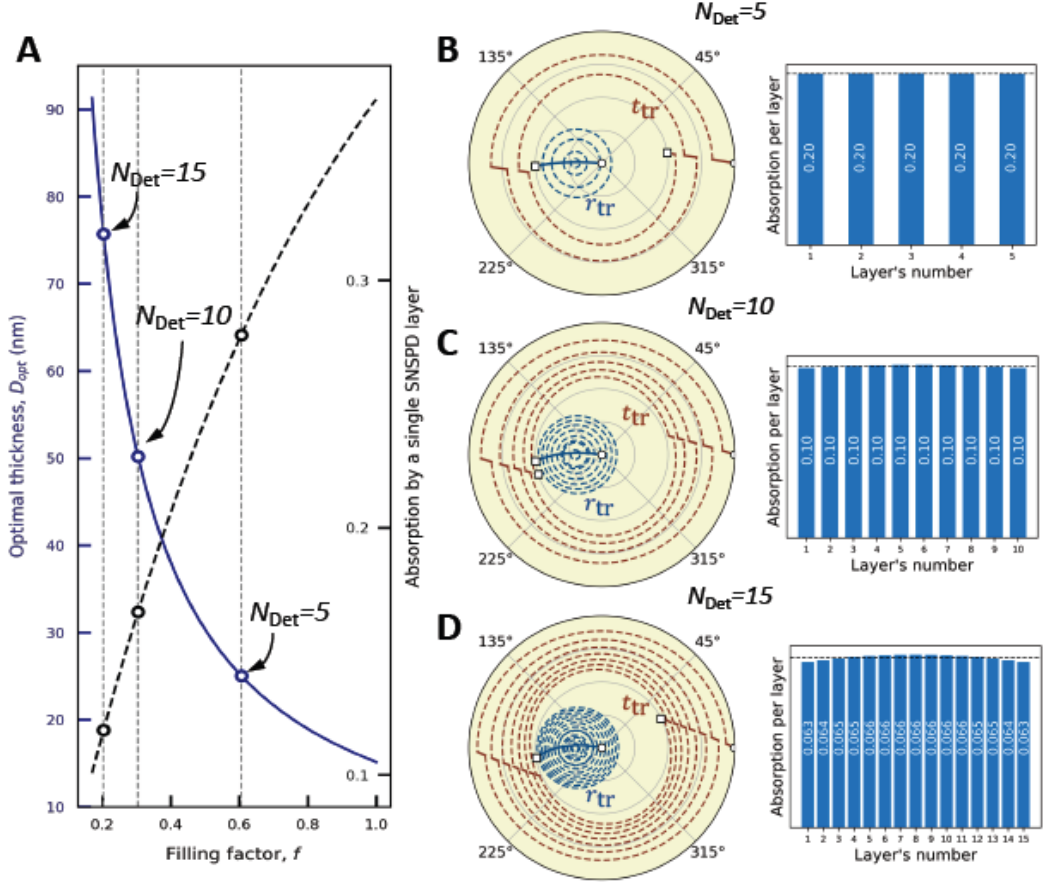

**Figure 5.** Performance of a multi-layer distributed detector. A, left vertical axis: Optimal thickness (solid purple line) of a single-layer SNSPD as a function of the filling factor. A, right vertical axis: absorption of a 5 nm thick nanowire as a function of the filling factor (dashed black line). Empty circles mark parameters for fabricating five-, ten-, and fifteen-layer distributed detectors. B-D: For these three cases, circular diagrams and absorption per layer (under a standing wave illumination) are shown.

The number of sub-layers and their individual absorption are tightly interlinked with the absorption uniformity achievable throughout the detectors. Since the optical thickness of each constituting absorber is finite, absorption is not uniformly distributed. Non-uniform absorption would then result in some detectors having higher multi-photon absorption probability, which effectively reduces the photon number resolution of the array.

To quantify the uniformity of light absorption, we introduce the absorption non-uniformity parameter Δ as the standard deviation of the absorption per layer from uniform absorption:

$$\Delta = \sqrt{\frac{1}{N_{\text{Det}}} \sum_{i=1}^{N_{\text{Det}}} \left( A_{\text{coh},i} - \frac{A_{\text{coh}}}{N_{\text{Det}}} \right)^2},$$

where $A_{\text{coh}}$ and $A_{\text{coh},i}$ are the total coherent absorption of the distributed detector and coherent absorption of the $i^{th}$ layer, respectively, and $N_{Det}$ is the number of the detecting layers in the structure. By dividing Δ by $\Delta_{\max} = \frac{A_{\text{coh}}}{N_{\text{Det}}}\sqrt{N_{\text{Det}} - 1}$ which holds when all the absorption happens in one constituent detector (maximum absorption non-uniformity), one can define the normalized non-uniformity parameter:

$$\Delta_{\text{norm}} \equiv \frac{\Delta}{\Delta_{\max}} = \sqrt{\frac{\sum_{i=1}^{N_{\text{Det}}}\left(1 - \frac{A_{\text{coh},i}}{A_{\text{coh}}} N_{\text{Det}}\right)^2}{N_{\text{Det}}(N_{\text{Det}} - 1)}},$$

The parameter $\Delta_{\text{norm}}$ varies between zero for uniform absorption, $A_{\text{coh},i} = A_{\text{coh}}/N_{\text{Det}}$, and unity for maximally non-uniform absorption.

Figs. 5B-D show the circular diagrams and absorption per layer (under a standing wave illumination) of representative distributed detector structures with the number of sublayers $N_{\text{Det}}$=5, 10 and 15. The circular diagram for the five-layer detector, Fig. 5B (left), shows that the entire structure fulfils the required optical response of $t_{\text{tr}} \approx 0.5$ and $r_{\text{tr}} \approx -0.5$ with conventional 5 nm thick nanowire detectors with 0.61 filling factor. The absorption uniformity of such structure is shown in Fig. 5B (right). As the total thickness of five sublayers (25 nm) is well within the subwavelength approximation, the absorption is uniform, with negligible non-uniformity parameter $\Delta_{\text{norm}}$=0.0008. Increasing the number of sublayers to ten requires a reduction of the filling factor of each nanowire absorber to 0.30 (Fig. 5A) to achieve the required optical response, Fig. 5C (left). Absorption remains rather uniform throughout the ten sublayers, with $\Delta_{\text{norm}} = 0.0022$, Fig. 5C (right). Note that, for any distributed detector with an even number of layers, transmission and reflection coefficients are in phase, $t_{\text{tr}} \approx -0.5$ and $r_{\text{tr}} \approx -0.5$. Such a detector absorbs out-of-phase counter-propagating beams or $\lambda/4$-shifted standing wave [48,49,51]. For a fifteen-layer distributed detector composed of nanowire sublayers with a filling factor of 0.20 (Fig. 5A), deviation from the subwavelength approximation is still tolerable. While the phase difference between $t_{\text{tr}}$ and $r_{\text{tr}}$ slightly decreases, Fig. 5D (left), the total coherent absorption reaches 98% with uniform absorption per layer, $\Delta_{\text{norm}} = 0.0040$, Fig. 5D (right).

To increase the number of layers even further, an additional optimization would be required where the more general condition $t_{\text{tr}} \approx \pm r_{\text{tr}} < 0.5$ can be reached with detectors of larger total thickness. The latter condition still guarantees perfect absorption but not perfect transmission. At the same time, a distributed detector can be used as a unit cell for spatial or temporal multiplexing, increasing the total number of detectors by order of magnitude. In the former case, a 3D detector can be assembled with multiplication in all three spatial coordinates.

In addition to PNR capability, a multi-layered distributed structure can substantially improve the temporal performance of single-photon detection in two ways. First, constituent nanowire sublayers have a lower filling factor and, thus, shorter nanowires. This, in turn, decreases the kinetic inductance of the nanowire, shortening the detectors' recovery time (recovery time is inversely proportional to the kinetic inductance). Second, multiple sublayers prevent two consecutive photons from being absorbed within the same sublayer allowing higher photon flux to be registered.

**Photon-number resolution with a distributed detector**

In this section, we first reproduce the framework for analyzing PNR detection in conventional (incoherent) schemes, like spatial and temporal multiplexing, following Refs. [62,63]. Then, we apply this procedure to coherent detection with a distributed detector and analyze in detail the performance of the ten-layer distributed SNSPD introduced in the previous section.

According to the postulate of quantum mechanics, the measurement procedure is described by a set of measurement operators and a corresponding transformation of the quantum state after the measurement. When the latter is of no interest, as in the case of light detection where photons are absorbed, the POVM ("positive operator-valued measure") formalism is widely used [64]. For a single on-off detector, like SNSPD, the POVM set (set of operators describing the measurement process) consists of two operators [62]

$$\hat{\pi}_0 = |0\rangle\langle 0| =: e^{-\hat{n}}: \quad \text{and} \quad \hat{\pi}_1 = \hat{1} - \hat{\pi}_0,$$

which are associated with the corresponding outcomes of the measurement: zero and one count. Here $|0\rangle$ is the vacuum state and $\hat{n}$ is number operator for the input light. The term $:e^{-\hat{n}}:$ originates from the photocounting formula (true photon statistics), and the symbol ": ... :" stands for the normal ordering of operators. The POVM set is complete, $\hat{\pi}_0 + \hat{\pi}_1 = \hat{1}$, which is required for probabilities of getting zero, $p(0) = \langle\hat{\pi}_0\rangle$, and one, $p(1) = \langle\hat{\pi}_1\rangle$, counts to sum up to unity. Here, $\langle...\rangle$ denotes averaging over the quantum state of the input light. For a realistic detector with the efficiency $\eta$ (ignoring dark counts), these probabilities are modified as

$$p(0) = \langle:e^{-\eta\hat{n}}:\rangle \text{ and } p(1) = \langle:\hat{1} - e^{-\eta\hat{n}}:\rangle. \tag{1}$$

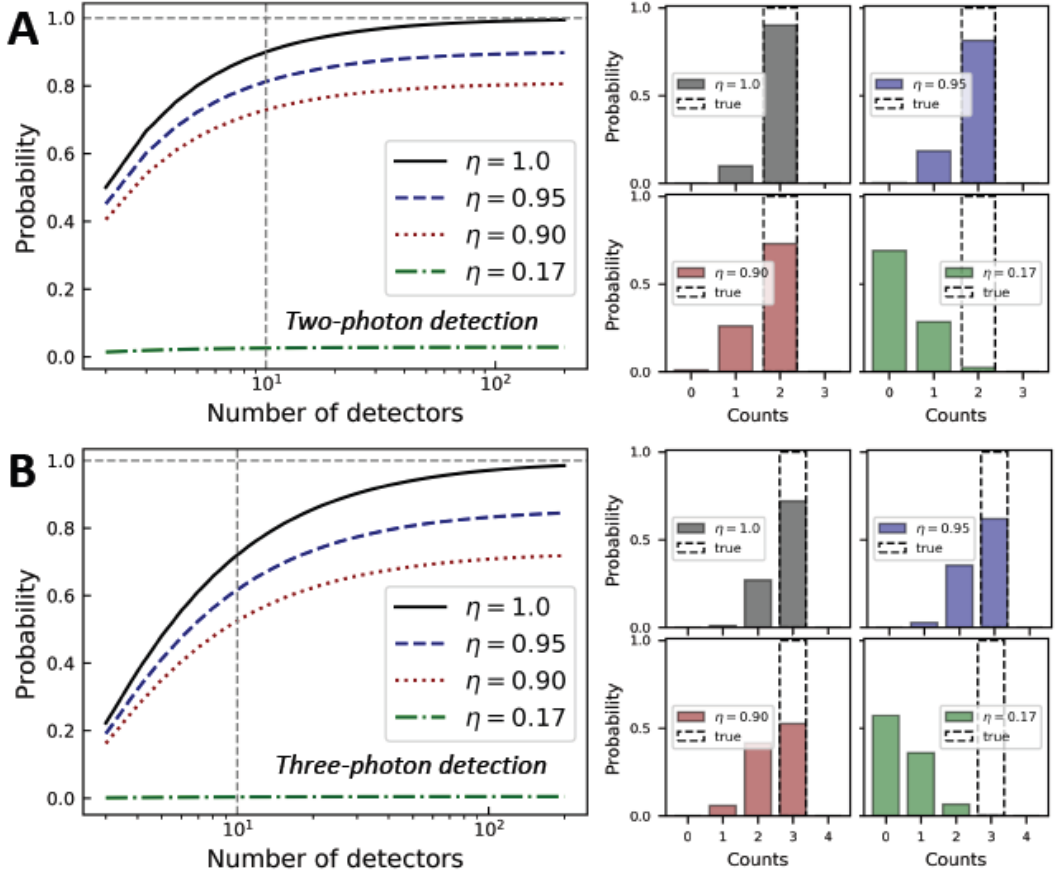

**Figure 6.** Photon number resolution with coherent and incoherent detection. (A, left): The probability of getting two clicks assuming detection of a two-photon number state as a function of the number of constituent detectors. Efficiency of each constituent detector is 100% (solid black line), 95% (dashed blue line), 90% (dotted red line), and 17% (dash-dotted green line). (A, right): The distribution of detection clicks for these four scenarios. (B): Same but for detection of a three-photon number state. For conventional incoherent schemes: the probability of correct photon number measurement drops exponentially with the efficiency of constituent detectors. For a coherent scheme with a distributed detector: efficient photon number resolution can be implemented with inefficient constituent detectors.

For spatial multiplexing, where *N* such on-off detectors monitor the input light, the probability of getting *k* counts is [62]

$$P(k) = \left\langle :\frac{N!}{k!(N-k)!}\left(e^{-\eta\frac{\hat{n}}{N}}\right)^{N-k}\left(\hat{1} - e^{-\eta\frac{\hat{n}}{N}}\right)^k: \right\rangle, \tag{2}$$

where uniform illumination is assumed. The same formula holds for the detection of *N* pulses in temporal multiplexing. In the Fock basis, Eq. (2) can be written (for $\eta = 1$) as [62]

$$P(k) = \langle \sum_{n=k}^{\infty} \frac{N!}{(N-k)!} \frac{1}{N^n} \left( \sum_{j=0}^{k} \frac{(-1)^j (k-j)^n}{j!(k-j)!} \right) |n\rangle\langle n|.  \tag{3}$$

By averaging over the input state of light, expressed in a Fock basis as well, this formula allows to find the photocounting statistics.

Alternatively, the detector can be described by a set of conditional probabilities $P(k|m)$ of getting $k$ counts given $m$ input photons. Again, by using the Fock state representation, the photocounting statistics can be derived for any state from these probabilities. The expression for $P(k|m)$ for a set of on-off detectors was derived in Ref. [63] as:

$$P(k|m) = \frac{1}{N^m} \binom{N}{k} \sum_{l=0}^{k} (-1)^l \binom{k}{l} [N - (N-k+l)\eta]^m,  \tag{4}$$

where the effect of non-ideal efficiency of constituent detectors is accounted for. Equation (4) is equivalent to (2) and (3) if the number state $|m\rangle$ is used for quantum averaging in (2) and (3).

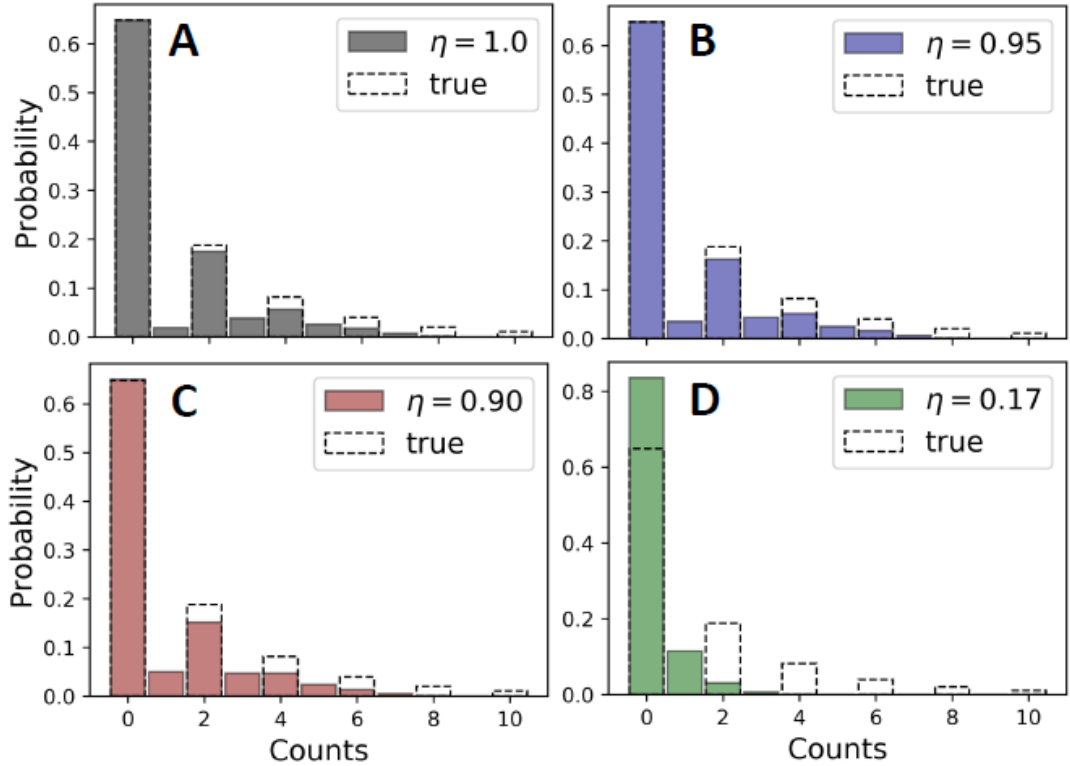

**Figure 7.** Measurement of photon statistics of the vacuum squeezed state ($\xi = 1$). Bar charts show the distribution of counts for incoherent detection with ten constituent detectors with the efficiency of 100% (A), 95% (B), 90% (C), and 17% (D). Dashed bars represent the true distribution. For incoherent detection with 17% efficient constituent detectors (D), the true distribution (dashed columns) is completely wiped in the measured distribution (green). For coherent detection with the same constituent detectors, the gray distribution in (A), close to the true distribution, holds.

The above formalisms are valid for the distributed detector as well, with one exception. As discussed earlier, the coherent detection scheme guarantees perfect and uniform absorption of the incoming light, irrespective of the absorption of constituent detectors. Therefore, while in Eqs. (2)-(4) the constituent detectors are described by (1) with $\eta < 1$ (or even $\ll 1$), for the entire distributed detector $\eta = 1$. In Fig. 6, we compare the performance of coherent and incoherent detection schemes at detecting two- and three-photon number states. The solid black line shows the probability of getting the correct measurement – two, Fig. 6A, and three, Fig. 6B, counts for the distributed detector. This result also holds for spatial and temporal multiplexing with 100% efficient detectors. The vertical gray dashed line corresponds to the detector composed of ten constituent detectors. The probability of getting the correct measurement is 90% and 74% for two and three-photon detection, respectively. The probability drops to 81% and 64% (69% and 54%) for the incoherent detection with the efficiency of each constituent detector of 95% (dashed blue line) and 90% (dotted red line), accordingly. As the efficiency of each constituent detector in the ten-layer distributed detector is just 17%, we also show the performance of the incoherent scheme composed of such constituent detectors (dash-dotted green line). The probability of correct detection, in this case, is just about 2.6% for a two-photon and 0.4% for a three-photon detection. For all described detection scenarios, we plot counts distribution on the right-hand side of Figs. 6A and 6B. The difference between the upper left (gray) and bottom right (green) distributions, corresponding to detection with the same constituent detectors but arranged according to coherent and incoherent detection schemes, clearly demonstrates the advantage of coherent detection over conventional multiplexing schemes when the efficiency of individual detectors is not unitary.

Another test for the efficiency of photon number resolution is the observation of non-classicality in photon number distributions for quantum states of light, such as the squeezed vacuum state. The true distribution is,

$$P_{sq}(n) = \begin{cases} \frac{1}{\cosh \xi} \left(\frac{\tanh \xi}{2}\right)^2 \frac{(n)!}{[(n/2)!]^2}, \text{ for even } n, \\ 0, \text{ for odd } n, \end{cases}$$

where $\xi$ is the squeezing parameter and $n$ is the number of photons. Accounting for the detectors' response (4), we evaluate count distribution, Fig. 7, for the four scenarios discussed above assuming $\xi = 1$. The non-classicality of the state is apparent for the ten-layer distributed detector, Fig. 7A, especially for low-count events: probabilities of one (1.9%) and three (3.8%) counts are noticeably lower than probabilities of two (17.4%) and four (5.6%) counts which are close to the true distribution. For the incoherent detection schemes with 95% and 90% efficient constituent detectors, Fig. 7B and Fig. 7C, the distribution blurs, hiding the non-classical features of the squeezed state's photocounting statistics. Finally, for the incoherent detection with 17% efficient detectors, the true distribution completely vanishes, and approaches the distribution of the Glauber coherent (classical) state.

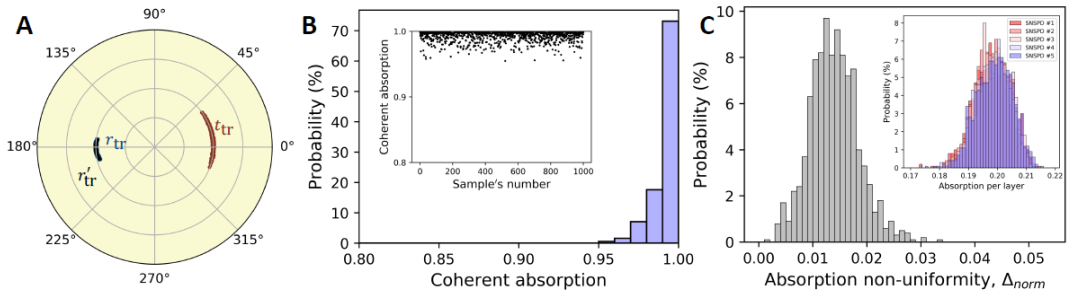

**Figure 8.** Statistical simulation of fabrication imperfection for the five-layer distributed detector. An ensemble of one thousand samples is generated where, for each sample, the thickness of nanowire sublayers and spacers is chosen randomly with fluctuation ±5% around the nominal values. A: Optical response under traveling wave illumination within the ensemble. Due to the asymmetry of the structure, the reflection coefficient is slightly different for illumination from the left, $r_{tr}$ (blue dots), and right, $r'_{tr}$ (black dots), sides. B: Fluctuation of coherent absorption in the ensemble. The inset shows the coherent absorption of each sample. C: Fluctuation of the absorption non-uniformity in the ensemble. The inset shows the fluctuation of the absorption per each nanowire sublayer in the ensemble.

**Robustness against fabrication imperfections**

Importantly, a multilayered distributed detector is robust against fabrication imperfections. Let us illustrate in detail the case of a five-layer detector. We ran a statistical (Monte-Carlo) simulation for the ensemble of one thousand distributed detectors (samples), allowing for the thickness of each layer

– nanowire and spacer, to randomly fluctuate within a $\pm 5\%$ interval around the nominal values. As shown in Fig. 8, in spite of the realistic variation in layers thickness, the amplitude transmission and reflection coefficients are still grouped around the optimal values of equal magnitudes and opposite phases, Fig. 8A. As a result, almost three-quarters of the samples still show complete absorption (within 99% to 100%), and all of them absorb more than 95% of light, Fig. 8B. The absorption non-uniformity parameter $\Delta_{\text{norm}}$ varies insignificantly among the samples but is less than 0.03 in most of the cases, Fig. 8C, revealing an anti-correlation between the absorption of the different nanowire sublayers: the decrease in absorption of one (or few) of the sublayers is compensated by the increase in absorption of other sublayers. Absorption of each sublayer falls in a narrow distribution around the nominal value, as shown in the inset of Fig. 8C.

**Generalized requirements for distributed absorbers**

In previous sections, we built distributed detectors by starting with a single-layer absorber of the optimal thickness taking into account superconducting material and nanowire design parameters. This single-layer absorber was "cut" into multiple sublayers placed at different anti-nodes of the standing wave. Under these assumptions, the standing wave picture guarantees that the distributed detector retains the optical response of the original layer with optimal thickness, as well as uniformity of absorption. In general, for a distributed detector with M-sublayers of any nature operating in the counter-propagating geometry, the amplitude transmission (*t*), reflection (*r*), and intensity absorption (*A*) coefficients of each sublayer should satisfy the following conditions (Supplementary Materials):

$$t = \frac{M}{M+1}, \qquad r = -\frac{1}{M+1}, \quad \text{and } A = \frac{2M}{(M+1)^2}. \tag{5}$$

Here, the sublayers are assumed to be of subwavelength thickness and spaced by $\lambda/2$-spacers. The optical response (5) can be achieved by proper design or adjusted empirically. This result is a generalization of what was presented above for multi-layered SNSPD detectors. For instance, according to (5), each constituent detector of a five-layer distributed detector should absorb 28% of light under traveling wave illumination. This is precisely the value we estimated for the five-layer SNSPD detector (Fig. 5A).

Similarly, the constituent sublayers of a *K*-layer distributed detector operating in the Salisbury screen geometry should possess the following amplitude transmission ($t'$), reflection ($r'$), and intensity absorption ($A'$) coefficients (Supplementary Materials):

$$t' = \frac{2K}{2K+1}, \qquad r' = -\frac{1}{2K+1}, \quad \text{and } A' = \frac{4K}{(2K+1)^2}. \tag{6}$$

Comparison of (5) and (6) reveals that, for distributed detectors with the same number of sublayers, *K=M*, sublayers should be less absorptive in the Salisbury screen geometry than in the counter-propagating configuration. Equivalently, given a certain optical response of the constituent detectors, half as many detectors are required in the counter-propagating geometry, *K=M/2*.

**DISCUSSION and CONCLUSIONS**

The coherent detection scheme proposed here has clear advantages over incoherent temporal multiplexing schemes as it does not delay acquisition beyond the intrinsic time-resolution of the constituent detectors. In fact, by relieving requirements on absorption, coherent detection generally allows for shorter, therefore faster SNSPDs. With respect to spatial multiplexing schemes, performance of coherent PNR detectors should be discussed in the context of their applications.

"Single-shot" applications like certain linear optical quantum computing and quantum communication protocols, rely on identification of the number of photons in *every* input optical pulse to determine success of the quantum computing trial or safety and reliability of the communication [1-4]. Such applications require PNR detectors that can efficiently discriminate between one- and multi-photon states within a single shot. For spatially multiplexed incoherent detection, a 10x10 matrix made of SNSPDs with efficiency of 90-95% would yield 80-90% 2-photon state resolution probability, Fig. 6A. If one combined the coherent detection scheme with multiplexed PNRs, a 2-layer 7x7 matrix or a 3-layer 6x6 matrix with approximately the same total number of the same detectors could readily yield 100% 2-photon state resolution probability if the absorption of the constituent detector arrays is properly adjusted (e.g., by varying the filling factor of the nanowire meander).

On the other hand, applications like quantum light source characterization or heralded quantum states preparation may operate in the "acquisition" regime: even if the PNR detectors are not very efficient, it is possible to wait a certain period of time to acquire statistics for source characterization or to realize a particular outcome of detection for state preparation [5,6,65]. The waiting time, however, scales exponentially with the detector's efficiency. While increasing efficiency of spatially multiplexed PNRs may be challenging due to detector proximity and thermal crosstalk, stacking of the same SNSPDs in a multilayer structure for coherent detection could become viable.

In conclusion, we have demonstrated an alternative method of photon number resolution which combines a coherent detection scheme and distributed arrangement of single-photon detectors. In contrast to conventional schemes operating by multiplexing optical modes in space or time, our method is based on the interaction of constituent detectors with a single optical mode. This results in perfect and uniform absorption of the incoming light, both of which are crucial for PNR detection. The coherent detection can be designed in a phase-sensitive counter-propagating geometry, where the absorption level can be controlled from 0 to 100%. This design can be of interest for feed-forward protocols or self-configuring optical networks. When this functionality is not required, the phase-insensitive Salisbury screen design can be used instead, where a distributed detector always operates in the perfect absorption regime. Given the maturity of multilayer deposition technology, the fabrication of distributed detectors is feasible and can be performed with high precision. We note that beyond quantum light detection, coherent schemes provide benefits for other protocols of quantum optics as well, including quantum memory [66] and deterministic entanglement generation in multi-nodal quantum networks [51].

Finally, we would like to note that while we based our analysis on thin detectors in the subwavelength approximation, which is well suited to describe SNSPDs, transition edge sensors and superconducting microbridges, it is straightforward to generalize the analysis to any constituent detectors of arbitrary thickness, including single-photon avalanche diodes. Thus, coherent detection with distributed absorption provides a general method for robust and efficient PNR detection in a variety of quantum technology platforms.

**Methods**

The transfer matrix method is widely used to calculate the optical response of multilayer structures assuming single-side illumination. The transfer matrix matches the input and output fields of the structure,

$$\begin{pmatrix} A_0 \\ B_0 \end{pmatrix} = M \begin{pmatrix} A_{sub} \\ B_{sub} \end{pmatrix},$$

where $A_0$, $B_0$ and $A_{sub}$ are amplitudes of input, reflected and transmitted waves, respectively. The amplitude $B_{sub}$ is set to zero as non-physical for single-side illumination:

$$\begin{pmatrix} A_0 \\ B_0 \end{pmatrix} = M \begin{pmatrix} A_{sub} \\ 0 \end{pmatrix}. \tag{7}$$

The transfer matrix $M$ results from the consecutive multiplication of matrices describing the transformation of the fields on interfaces and matrices describing propagation within layers [67]. For illumination from the opposite side:

$$\begin{pmatrix} 0 \\ B'_0 \end{pmatrix} = M_{total} \begin{pmatrix} A'_{sub} \\ B'_{sub} \end{pmatrix}, \tag{8}$$

where $A'_{sub}$, $B'_{sub}$ and $B'_0$ are amplitudes of input, reflected and transmitted waves, respectively. Thus the coherent illumination of the structure is a sum of (7) and (8), where the phase difference between the input fields should be accounted for. By decomposing the transfer matrix on the constituent matrices, the amplitudes of the fields can be restored at any point within the structure, and accordingly, the absorption of each layer can be evaluated.

**ACKNOWLEDGEMENTS**

This work was supported by the Quantum Engineering Programme of the National Research Foundation (NRF-QEP1 and NRF2021-QEP2-01-P01) and by the Tier 3 Programme of the Ministry of Education (MOE2016-T3-1-006 (S)) of Singapore.

**DATA AVAILABILITY**

Concepts discussed in this manuscript were first disclosed in "Method for photon number resolving detection without optical mode multiplication", Singapore provisional patent application No. 10202105111Q (filed on 17 May 2021), PCT Patent Application No. PCT/SG2022/050318 filed on 13 May 2022, claiming priority to Singapore provisional patent application number 0202105111Q. The data that support the findings of this study are openly available in NTU research data repository DR-NTU (data) at https:// doi.org/XXXXXXXX.